\documentclass[final,12pt,times]{elsarticle}
\usepackage{amssymb}
\usepackage{color}

\journal{Materials Today Communications}

\begin{document}

\begin{frontmatter}



\title{Superconducting properties of eutectic high-entropy alloy superconductor NbScTiZr}


\author[inst1]{Jiro Kitagawa}
\ead{j-kitagawa@fit.ac.jp}
\affiliation[inst1]{organization={Department of Electrical Engineering, Faculty of Engineering, Fukuoka Institute of Technology},%Department and Organization
            addressline={3-30-1 Wajiro-higashi, Higashi-ku}, 
            city={Fukuoka},
            postcode={811-0295},
            country={Japan}}

\author[inst2]{Haruto Ueta}
\affiliation[inst2]{organization={Department of Electrical Engineering, Faculty of Science and Engineering, Kyushu Sangyo University},%Department and Organization
            addressline={2-3-1 Matsukadai, Higashi-ku}, 
            city={Fukuoka},
            postcode={813-8503},
            country={Japan}}

\author[inst3]{Yuto Watanabe}

\affiliation[inst3]{organization={Department of Physics, Tokyo Metropolitan University},%Department and Organization
            city={Hachioji},
            postcode={192-0397}, 
            country={Japan}}

\author[inst1]{Takeru Seki}

\author[inst3]{Yoshikazu Mizuguchi}

\author[inst2]{Terukazu Nishizaki}

\begin{abstract}
The influence of annealing on the superconducting properties of eutectic high-entropy alloy NbScTiZr was examined by measuring magnetization, electrical resistivity, and zero-field specific heat.
Additionally, the extent of lattice strain was assessed via the analysis of lattice parameters and Vickers microhardness. 
The greatest lattice strain was inferred in the sample annealed at 400 $^{\circ}$C. 
Field-dependent magnetization datasets indicate enhanced flux pinning in the as-cast, 400 $^{\circ}$C annealed, and 600 $^{\circ}$C annealed samples.
The superconducting parameters such as upper critical field, Ginzburg-Landau coherence length, and magnetic penetration depth do not strongly depend on the annealing temperature.
However, the Maki parameter shows a peak at 400 $^{\circ}$C annealing temperature and correlates with the magnitude of lattice strain. 
This suggests that greater lattice strain may enhance the Maki parameter by altering the orbital-limited field through the modification of flux pinning strength.
Although the influence of lattice strain is less discernible in zero-field specific heat measurements, the superconducting parameters deduced from specific heat data suggest a potential for strong-coupled superconductivity in samples heat-treated above 600 $^{\circ}$C. 
Our investigation underscores the substantial impact of lattice strain on the Maki parameter of eutectic HEA superconductors, primarily through the modulation of flux pinning strength.
\end{abstract}


%\item HfMoNbTiZr is a new high-entropy alloy (HEA) superconductor.
%\item The superconducting properties are compared among HEA superconductors.
%\item The higher Debye temperature suppresses superconducting critical temperature $T_\mathrm{c}$.
%\item High hardness leads to lower $T_\mathrm{c}$.


\end{frontmatter}


\section{Introduction}
In traditional alloys, a minor quantity of additional elements is incorporated into a predominant element. 
The concept of high-entropy alloy (HEA), introduced in 2004, has disrupted the conventional paradigm of alloy design\cite{Cantor:MSEA2004,Yeh:AEM2004}. 
HEAs typically consist of four or more principal elements with approximately equal atomic proportions\cite{Senkov:IM2010,Zhang:JSA2024}. 
These metallic alloys have garnered considerable attention owing to their array of unconventional properties, including high-temperature strength, heightened resistance to radiation, exceptional thermoelectric figure of merit, improved magnetic refrigeration characteristics, good soft ferromagnetism, and interesting superconductivity\cite{Li:PMS2021,Miracle:Entropy2014,Li:JNM2019,Jiang:Science2021,Law:JMR2023,Nakamura:AIPAd2023,Chaudhary:MT2021,Kitagawa:JMMM2022,Kitagawa:Metals2020,Sun:PRM2019}.
Superconducting HEAs have emerged as a prominent research focus since the discovery of an HEA superconductor in 2014\cite{Kozelj:PRL2014}. 
Presently, HEA superconductivity is reported in diverse crystal structures such as body-centered cubic (bcc)\cite{Rohr:PRM2018,Marik:JALCOM2018,Ishizu:RINP2019,Harayama:JSNM2021,Sarkar:IM2022,Motla:PRB2022,Kitagawa:RHP2022,Kitagawa:JALCOM2022,Li:JPCC2023}, hexagonal close-packed (hcp)\cite{Lee:PhysicaC2019,Marik:PRM2019,Browne:JSSC2023}, face-centered cubic\cite{Zhu:JALCOM2022}, CsCl-type\cite{Stolze:ChemMater2018}, A15\cite{Wu:SCM2020,Yamashita:JALCOM2021}, NaCl-type\cite{Mizuguchi:JPSJ2019,Yamashita:DalTran2020}, $\alpha$ (or $\beta$)-Mn-type\cite{Stolze:JMCC2018,Xiao:SM2023}, $\sigma$-phase type\cite{Liu:ACS2020}, CuAl$_{2}$-type\cite{Kasen:SST2021}, BiS$_{2}$-based\cite{Sogabe:SSC2019}, and YBCO-based\cite{Shukunami:PhysicaC2020} structures. 
Noteworthy discoveries in the domain of HEA superconductors include the robustness of superconductivity against high pressures and irradiation, deviations from the Matthias rule, and the attainment of high critical current densities\cite{Guo:PNAC2017,Kasem:SR2022,Jung:NC2022,Hattori:JAMS2023,Gao:APL2022,Yamashita:JJAP2022,Seki:JSNM2023,Kim:JMST2024}. 
Additionally, the high hardness or shape memory effect exhibited by certain superconducting HEAs holds promise for developing multifunctional devices\cite{Kitagawa:JALCOM2022,Egilmez:PRM2021}.

The concept of eutectic HEA represents a recent innovation that has sparked a revolution in traditional HEAs\cite{Lu:SR2014}. 
Eutectic high-entropy alloys combine the benefits of both eutectic alloys and HEAs, showcasing excellent castability and superior mechanical properties. 
Notably, high strength and good ductility can be concurrently achieved by controlling the two phases in eutectic HEAs\cite{Lu:SM2020}. 
However, the superconductivity in eutectic HEAs remains relatively unexplored\cite{Krnel:Materials2022}.
Our recent investigation on eutectic superconducting HEA NbScTiZr has unveiled the modulation of the superconducting critical temperature ($T_\mathrm{c}$) through thermal annealing\cite{Seki:JSNM2023}. 
The eutectic HEA comprises the bcc and hcp phases, with the bcc phase undergoing the superconducting transition. 
We observed a systematic enhancement of $T_\mathrm{c}$ from 7.9 to 9 K with increasing thermal annealing temperatures up to 800 $^{\circ}$C. 
Thermal annealing above 400 $^{\circ}$C results in grain size coarsening. 
Moreover, analysis of lattice parameters suggests the presence of lattice strain, particularly at lower annealing temperatures. 
We have also demonstrated a remarkably high critical current density for the as-cast NbScTiZr. 
Nevertheless, the impact of thermal annealing on superconducting parameters, such as critical fields and coherence length in NbScTiZr, remains unexplored. 
Investigating this aspect is essential for comprehending the role of lattice strain associated with the eutectic structure in HEA superconductivity.

In this study, we estimated the fundamental superconducting parameters through magnetic, transport, and zero-field thermal measurements of NbScTiZr samples prepared under various annealing temperatures. 
Magnetization measurements provide the lower and upper critical fields, facilitating the extraction of the Ginzburg-Landau coherence length and the magnetic penetration depth. 
We discuss the annealing temperature dependences of these parameters and the influence of lattice strain on flux pinning strength. 
The electrical resistivity measurements as a function of magnetic field are also informative for discussing upper critical field.
Zero-field specific heat measurements provide insights into the bulk nature of the transition associated with chemical homogeneity. 
Utilizing specific heat data, we compare thermal parameters with those of the BCS model in the weak coupling limit. 
Furthermore, lattice strain is probed via Vickers microhardness analysis.

\begin{table*}
\caption{\label{tab:table1}%
Chemical compositions of bcc and hcp phases and lattice parameters of prepared samples.}
\begin{center}
\begin{tabular}{ccc}
\hline
sample    & composition   & lattice parameter (\AA)  \\
\hline
as-cast    & bcc:Nb$_{26.6(4)}$Sc$_{22.3(5)}$Ti$_{25.5(5)}$Zr$_{25.6(4)}$ & $a$=3.359(2) \\
           & hcp:Nb$_{21.3(4)}$Sc$_{30(1)}$Ti$_{23.0(7)}$Zr$_{25.8(5)}$  & $a$=3.262(3), $c$=5.152(3)   \\
400 $^{\circ}$C   & bcc:Nb$_{27.9(3)}$Sc$_{21.6(2)}$Ti$_{25.3(3)}$Zr$_{25.2(2)}$  & $a$=3.308(2) \\
           &hcp:Nb$_{22.1(5)}$Sc$_{32.3(9)}$Ti$_{21.3(5)}$Zr$_{24.2(2)}$ & $a$=3.252(1), $c$=5.131(1)   \\
600 $^{\circ}$C   & bcc:Nb$_{29.1(6)}$Sc$_{19.1(4)}$Ti$_{29.8(9)}$Zr$_{22(1)}$  & $a$=3.314(1) \\
           &hcp:Nb$_{18.0(5)}$Sc$_{32(1)}$Ti$_{19(1)}$Zr$_{30.8(7)}$ & $a$=3.254(1), $c$=5.133(2)   \\
800 $^{\circ}$C    & bcc:Nb$_{33(1)}$Sc$_{15(1)}$Ti$_{32(1)}$Zr$_{20(1)}$  & $a$=3.325(1)   \\
           &hcp:Nb$_{13(1)}$Sc$_{39.1(9)}$Ti$_{15.1(4)}$Zr$_{32.6(6)}$
       & $a$=3.252(2), $c$=5.138(3)   \\
1000 $^{\circ}$C    & bcc:Nb$_{39.0(8)}$Sc$_{4.2(6)}$Ti$_{38.0(6)}$Zr$_{18.9(8)}$ & $a$=3.360(1)   \\
           & hcp:Nb$_{2.6(7)}$Sc$_{56(1)}$Ti$_{7.3(7)}$Zr$_{34.1(5)}$ & $a$=3.260(1), $c$=5.141(2)    \\
\hline
\end{tabular}
\end{center}
\end{table*}

\section{Materials and Methods}
The as-cast sample was made employing a homemade arc furnace under an argon atmosphere. 
The constituent materials comprised Nb wire (Nilaco, 99.9 \%), Sc chips (Kojundo Chemical Laboratory, 99 \%), Ti wire (Nilaco, 99.9 \%), and Zr wire (Nilaco, 99.5 \%). 
The mixture of elements with the atomic ratio of Nb:Sc:Ti:Zr=1:1:1:1 underwent melting on a water-cooled Cu hearth. 
Subsequent annealing at 400 $^{\circ}$C, 600 $^{\circ}$C, 800 $^{\circ}$C, or 1000 $^{\circ}$C for 4 days was conducted in an electric furnace, with the sample enclosed within an evacuated quartz tube.

X-ray diffraction (XRD) patterns were acquired utilizing an X-ray diffractometer (Shimadzu XRD-7000L) employing Cu-K$\alpha$ radiation in the Bragg-Brentano configuration. 
Thin slabs cut from the samples were utilized because of the difficulty in obtaining fine powders due to high ductility.
The temperature-dependent dc magnetization $M$($T$) and isothermal magnetization curves were recorded employing a SQUID magnetometer (MPMS3, Quantum Design). 
The temperature dependences of electrical resistivity $\rho$($T$) under several external fields up to 9 T were measured by the four-probe method using PPMS equipment (Quantum Design). 
Specific heat $C$ was determined via the thermal-relaxation method employing PPMS equipment (Quantum Design). 
Vickers microhardness was assessed on polished surfaces under a 2.94 N load applied for 10 seconds using a Shimadzu HMV-2T microhardness tester.

\begin{figure}
\begin{center}
\includegraphics[width=1\linewidth]{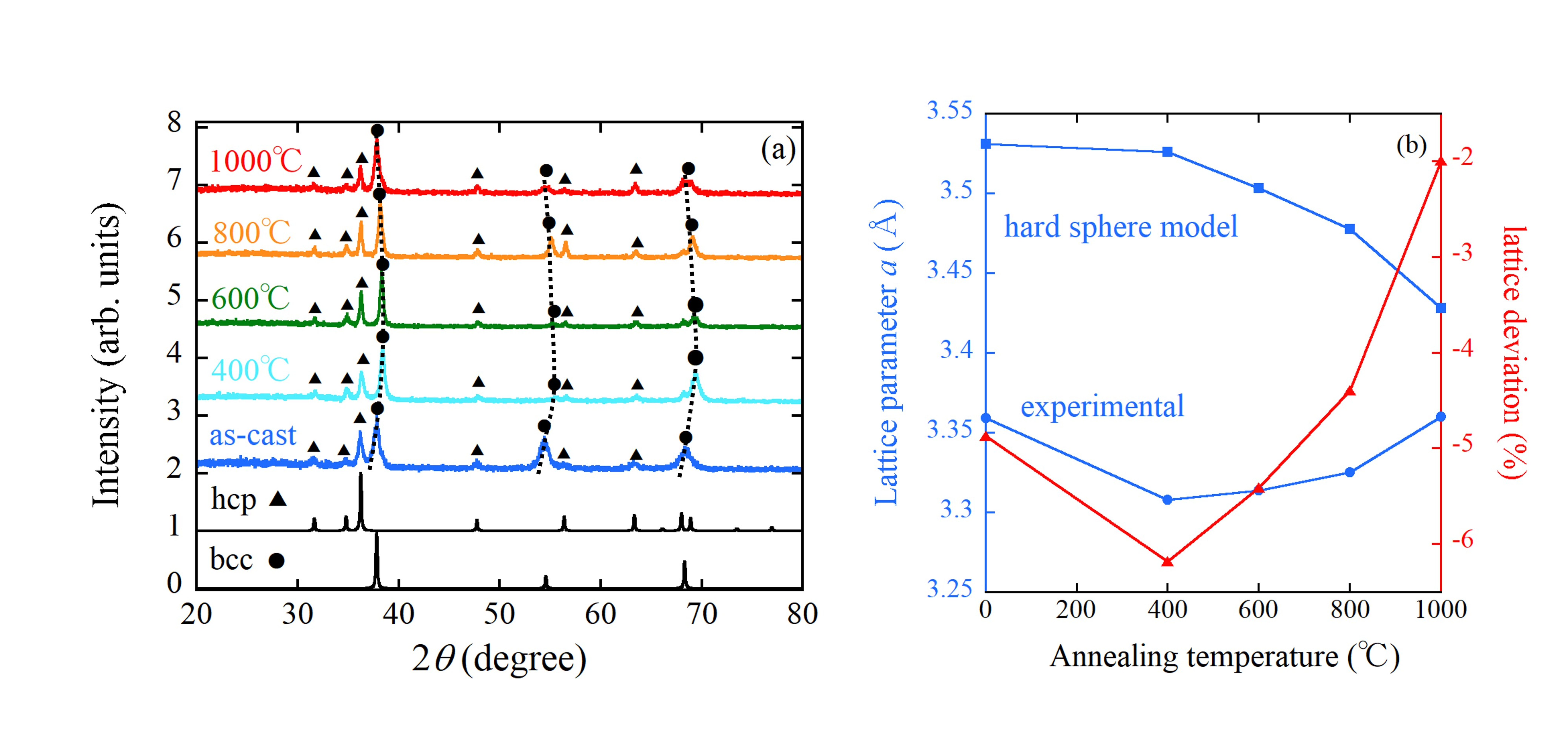}
\caption{\label{fig1}(a) XRD patterns of NbScTiZr samples prepared under various annealing conditions. The simulation patterns of the bcc ($a$=3.359 \AA) and hcp ($a$=3.262 \AA, $c$=5.152 \AA) structures are also presented. The origin of each XRD pattern is shifted by a value for the clarity. (b) Annealing temperature dependences of lattice parameter (filled blue circles) and that calculated using the hard-sphere model (filled blue squares). The annealing temperature dependence of lattice deviation, defined as $(a_\mathrm{exp}-a_\mathrm{hard})/a_\mathrm{hard}\times 100$, where  $a_\mathrm{exp}$ and $a_\mathrm{hard}$ are the lattice parameters of the experimental data and the hard-sphere model, respectively, is also shown.}
\end{center}
\end{figure}

\begin{figure}
\begin{center}
\includegraphics[width=1.1\linewidth]{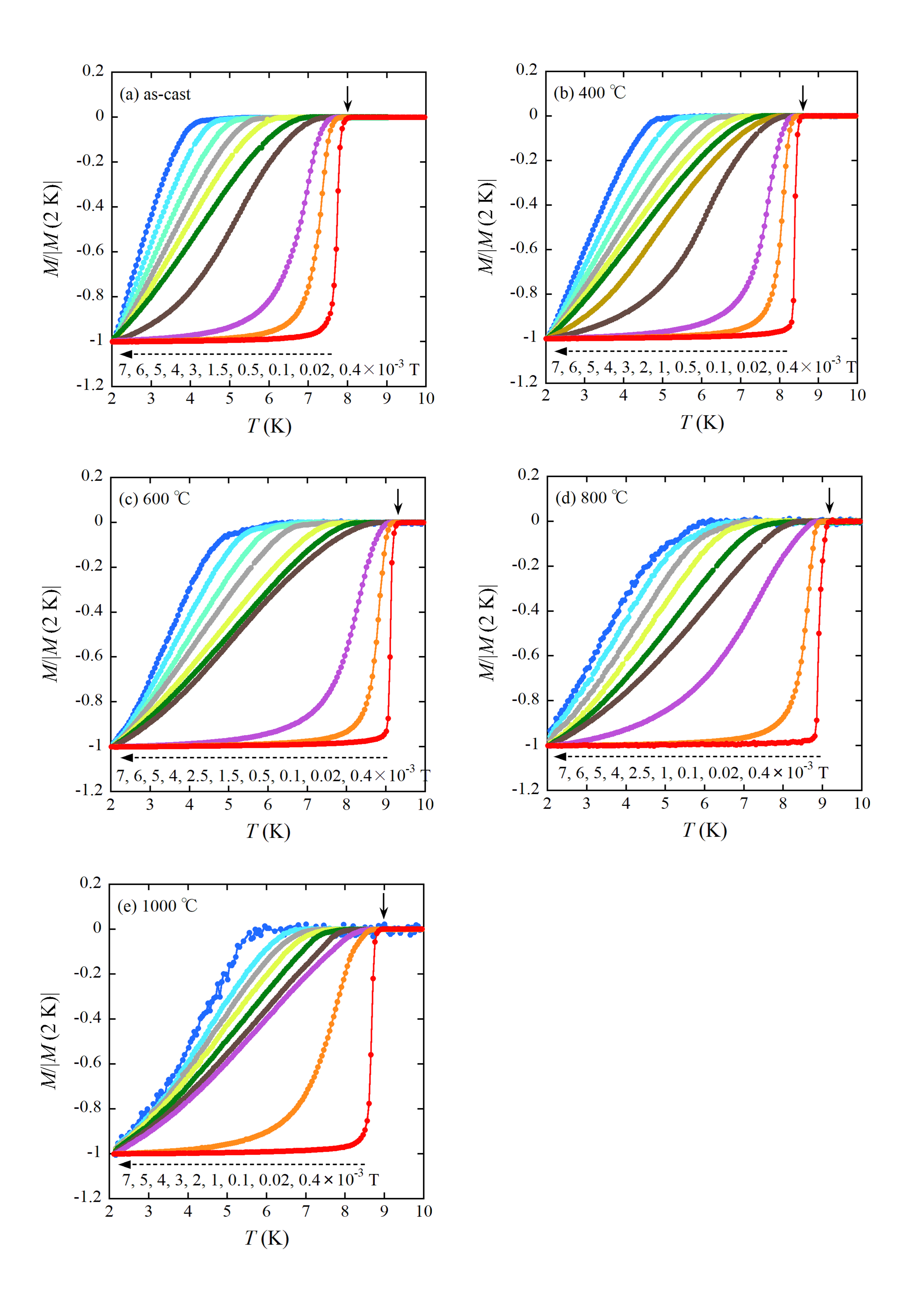}
\caption{\label{fig2} Temperature-dependent ZFC magnetization measured under external fields, as denoted in figure for (a) as-cast, (b) 400 $^{\circ}$C annealed, (c) 600 $^{\circ}$C annealed, (d) 800 $^{\circ}$C annealed, and (e) 1000 $^{\circ}$C annealed samples. $M$ is normalized by the absolute value of $M$ at 2.0 K.}
\end{center}
\end{figure}

\section{Results and Discussion}
Figure \ref{fig1}(a) illustrates the XRD patterns of both as-cast and thermally annealed samples. 
The XRD pattern of the as-cast sample manifests peaks corresponding to bcc and hcp phases, denoted by filled circles and triangles, respectively (also refer to the simulated patterns of bcc and hcp structures). 
The Bragg reflection peaks of the thermally annealed samples are likewise assignable to the bcc or hcp structure. 
Notably, the positions of the Bragg reflection peaks shift towards higher angles as the annealing temperature progresses from the as-cast state to 400 $^{\circ}$C annealing and subsequently towards lower angles with further annealing above 400 $^{\circ}$C (refer to the dotted curves marked for the bcc phase in Fig.\hspace{1mm}\ref{fig1}(a)). 
After assigning Miller indices, the lattice parameter of each phase is determined employing the least square method, as summarized in Table \ref{tab:table1}. 
The obtained parameters of the bcc phase are plotted against the annealing temperature in Fig.\hspace{1mm}\ref{fig1}(b), considering the as-cast state as equivalent to 0 $^{\circ}$C annealing. 
We check the influence of lattice strain by employing the hard-sphere model with no consideration of lattice strain. 
In this model, the ideal lattice parameter is estimated based on the chemical composition derived from energy-dispersive X-ray spectrometer coupled with field-emission scanning electron microscopy\cite{Seki:JSNM2023}. 
The chemical composition of the bcc phase, as reported in a previous study\cite{Seki:JSNM2023}, is provided in Table \ref{tab:table1} for each sample. 
The hard-sphere model predicts the lattice parameter expressed as 2.309$\bar{r}$, where $\bar{r}$ represents the composition-weighted atomic radius. 
By utilizing atomic radii of 1.429 \AA \hspace{1mm} for Nb, 1.641 \AA \hspace{1mm} for Sc, 1.4615 \AA \hspace{1mm} for Ti, and 1.6025 \AA \hspace{1mm} for Zr\cite{Miracle:AM2017}, the calculated lattice parameters are also depicted in Fig.\hspace{1mm}\ref{fig1}(b). 
Lattice strain is assessed by the disparity between the experimental lattice parameter and that derived from the hard-sphere model. 
This disparity is denoted as lattice deviation and quantified by $(a_\mathrm{exp}-a_\mathrm{hard})/a_\mathrm{hard}\times 100$, where $a_\mathrm{exp}$ and $a_\mathrm{hard}$ are the lattice parameters of the experimental data and the hard-sphere model, respectively. 
Figure \ref{fig1}(b) further illustrates the annealing temperature dependence of the lattice deviation. 
The deviation exhibits a shallow minimum at 400 $^{\circ}$C, reaching -6 \%. 
This value is relatively higher compared to single-phase HEA superconductors such as HfMoNbTiZr\cite{Kitagawa:JALCOM2022} (lattice deviation: -1.7 \%) and (Ti$_{35}$Hf$_{25}$)(Nb$_{25}$Ta$_{5}$)Re$_{10}$\cite{Hattori:JAMS2023} (lattice deviation: -2.3 \%). 
This outcome supports the manifestation of lattice strain, particularly in samples annealed at lower temperatures.
Additionally, we evaluated the crystallite size $D$ of the bcc phase utilizing the Scherrer equation $D=K\lambda/\beta_\mathrm{XRD}\mathrm{cos}\theta$, where $\beta_\mathrm{XRD}$ represents the full width at half maximum of the diffraction peak,  $\lambda$ denotes the X-ray wavelength, and $K$ signifies a numerical coefficient. 
Assuming $K$=0.9, the $D$ values computed from the principal peaks of all samples are 14 nm (as-cast), 17 nm (400 $^{\circ}$C), 20 nm (600 $^{\circ}$C), 18 nm (800 $^{\circ}$C), and 15 nm (1000 $^{\circ}$C), respectively.

\begin{figure}
\begin{center}
\includegraphics[width=1\linewidth]{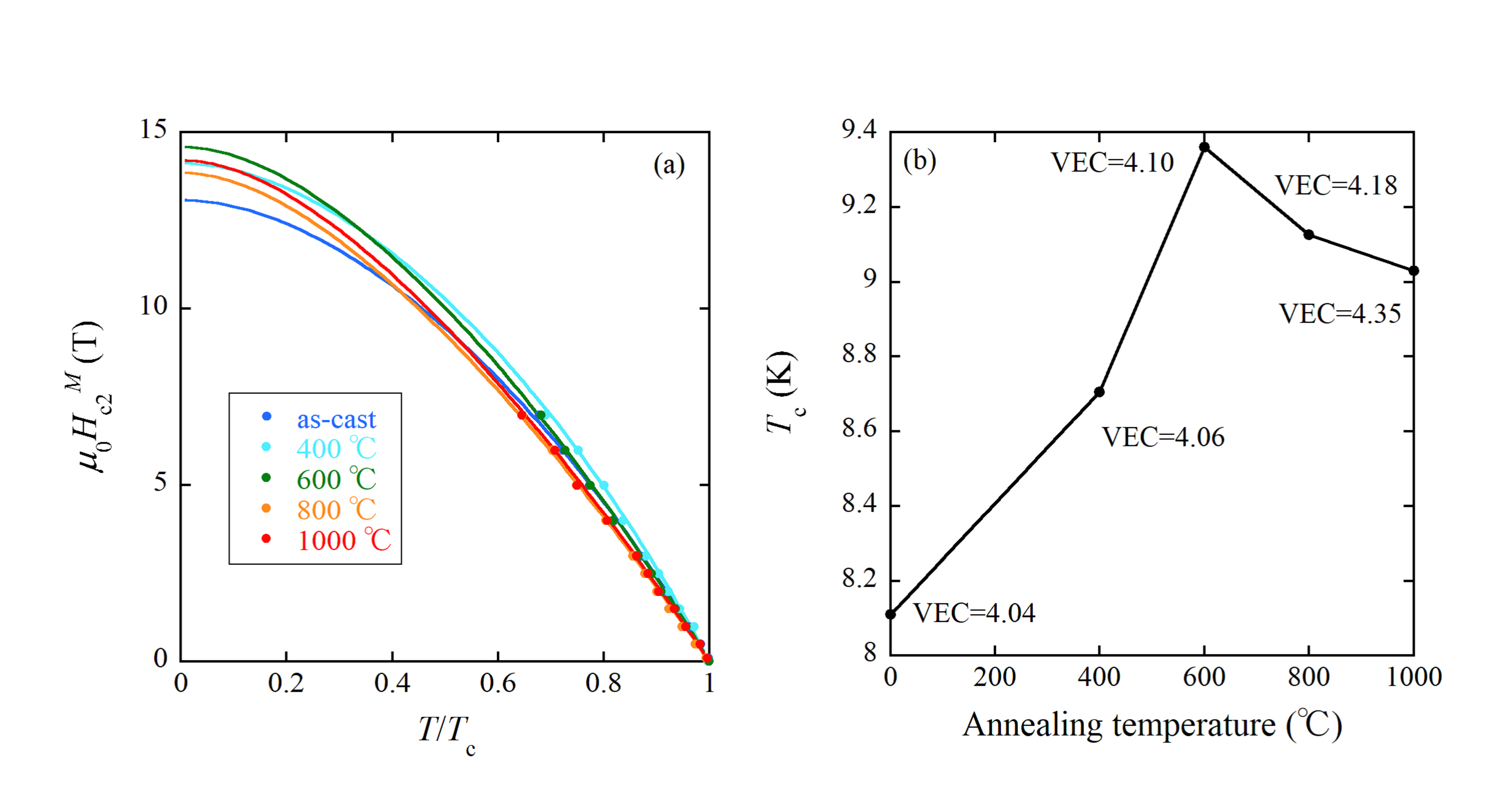}
\caption{\label{fig3} (a) Temperature dependences of upper critical field of NbScTiZr samples. The figure is derived from Figs.\hspace{1mm}\ref{fig2}(a)-(e). The solid curves show the fitting results using eq. (\ref{eq:hc2}). (b) Annealing temperature dependence of $T_\mathrm{c}$ extracted using eq. (\ref{eq:hc2}).}
\end{center}
\end{figure}

Figures \ref{fig2}(a)-(e) display the temperature dependences of magnetization measured under the zero-field cooled (ZFC) condition, with the external field $\mu_{0}H$ ranging from 0.4 mT to 7 T for all samples. 
Each $M$($T$) dataset is normalized by the absolute value of $M$ at 2 K. 
The onset temperature of diamagnetism at $\mu_{0}H$=0.4 mT mirrors $T_\mathrm{c}$ under zero magnetic field (also indicated by arrows in Figs.\hspace{1mm}\ref{fig2}(a)-(e)). 
As the annealing temperature increases to 600 $^{\circ}$C, $T_\mathrm{c}$ experiences enhancement. 
However, further increments in the annealing temperature marginally diminish $T_\mathrm{c}$. 
We determined the upper critical field $\mu_{0}H_\mathrm{c2}$ using the $M$($T$) datasets in Figs.\hspace{1mm}\ref{fig2}(a)-(e). 
The onset of the diamagnetic signal is defined as $T_\mathrm{c}$ at a $\mu_{0}H$ value (see also Fig.\hspace{1mm}S1 to S10 in Supplementary Material), enabling the construction of a $\mu_{0}H_\mathrm{c2}$ vs. temperature plot as illustrated in Fig.\hspace{1mm}\ref{fig3}(a). 
The temperature dependences of $\mu_{0}H_\mathrm{c2}$ are analyzed utilizing the Werthamer–Helfand–Hohenberg (WHH) model: 
\begin{equation}
\mathrm{ln}\frac{1}{t}=(\frac{1}{2}+\frac{i\lambda_\mathrm{SO}}{4\gamma})\psi\left(\frac{1}{2}+\frac{h+\lambda_\mathrm{SO}/2+i\gamma}{2t} \right)+(\frac{1}{2}-\frac{i\lambda_\mathrm{SO}}{4\gamma})\psi\left(\frac{1}{2}+\frac{h+\lambda_\mathrm{SO}/2-i\gamma}{2t} \right)-\psi\left(\frac{1}{2}\right)
\label{eq:hc2}
\end{equation}
, where $t=T/T_\mathrm{c}$, $\lambda_\mathrm{SO}$ is the spin-orbital scattering parameter, $\gamma=[(\alpha_\mathrm{M}h)^{2}-(\lambda_\mathrm{SO}/2)^{2}]^{1/2}$ ($\alpha_\mathrm{M}$: Maki parameter), and $\psi(x)$ is the digamma function.
In equation (\ref{eq:hc2}), $h$ is defined as 
\begin{equation}
h=\frac{4H_\mathrm{c2}(T)}{\pi^{2}T_\mathrm{c}(-dH_\mathrm{c2}(T)/dT)_{T=T\mathrm{c}}}.
\label{eq:h-def}
\end{equation}
For reliable fitting, we refer to the literature reporting the WHH model parameters of NbTi alloy\cite{Neuringer:PRL1966}.
The reported values of $\alpha_\mathrm{M}$ and $\lambda_\mathrm{SO}$ are 1.3 and 4.5, respectively, which are suitable for the 800 $^{\circ}$C and 1000 $^{\circ}$C annealed samples with weak lattice strain.
The same parameters can reproduce the temperature dependences of $\mu_{0}H_\mathrm{c2}$ of these samples, as shown in Fig.\hspace{1mm}\ref{fig3}(a). 
The obtained $\mu_{0}H_\mathrm{c2}$(0) and $\alpha_\mathrm{M}$ values of all samples are summarized in Table \ref{tab:table2}.
The values of $\lambda_\mathrm{SO}$ are 1.0, 1.0, 2.0, 4.5, and 4.5 for the as-cast, 400 $^{\circ}$C annealed, 600 $^{\circ}$C annealed, 800 $^{\circ}$C annealed, and 1000 $^{\circ}$C annealed samples, respectively. 
In Table \ref{tab:table2}, we use the notation $\mu_{0}H_\mathrm{c2}^{M}$(0) and $\alpha_\mathrm{M}^{M}$ for $\mu_{0}H_\mathrm{c2}$(0) and $\alpha_\mathrm{M}$, respectively.
The annealing temperature dependence of $\mu_{0}H_\mathrm{c2}$(0) or $\alpha_\mathrm{M}$ is extensively discussed below. 
The $T_\mathrm{c}$ values extracted from the analysis of $\mu_{0}H_\mathrm{c2}$ are plotted as a function of annealing temperature in Fig.\hspace{1mm}\ref{fig3}(b), elucidating the impact of annealing temperature on $T_\mathrm{c}$ as mentioned earlier.

\begin{figure}
\begin{center}
\includegraphics[width=1\linewidth]{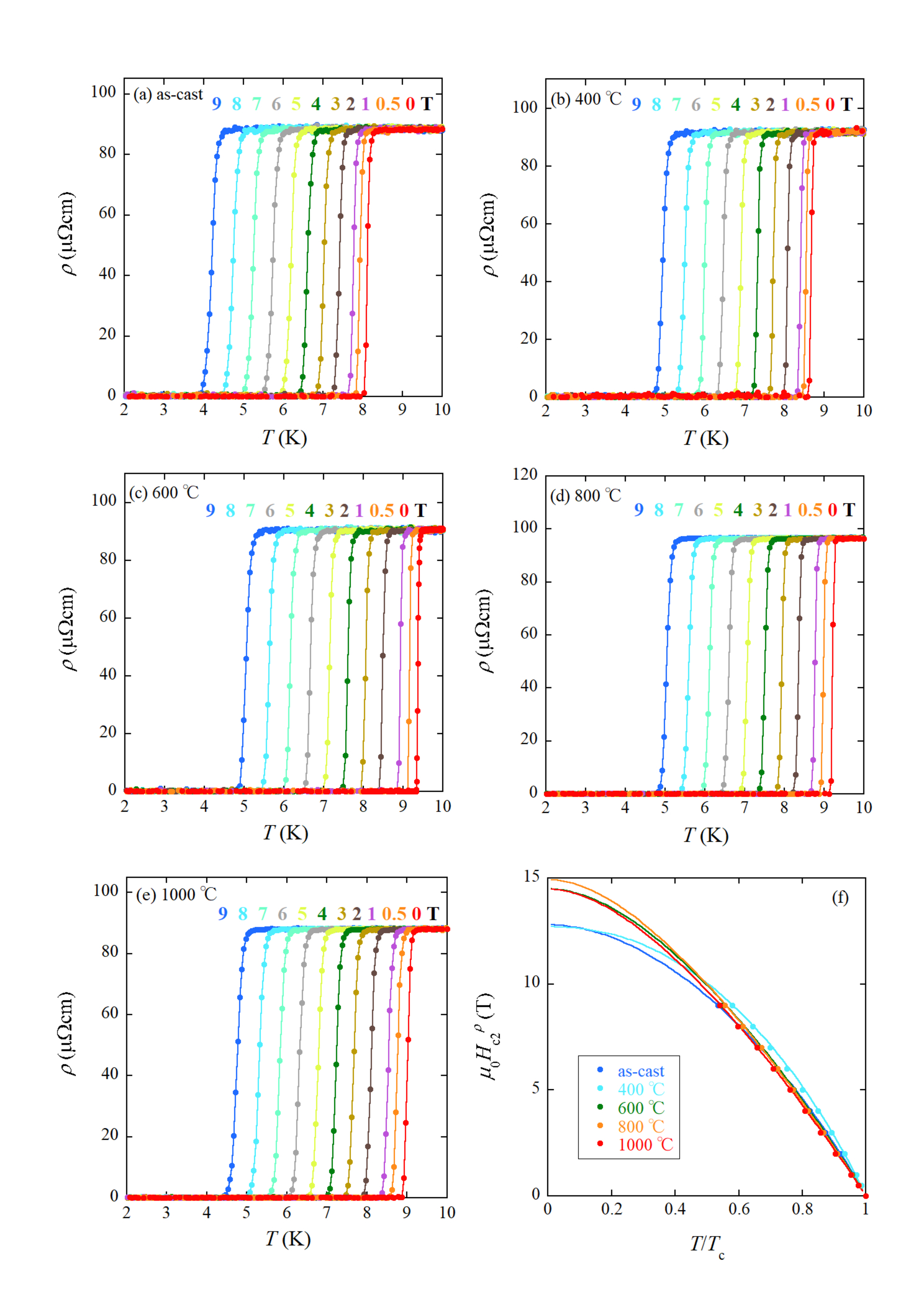}
\caption{\label{fig4} Temperature-dependent electrical resistivity measured under external fields, as denoted in figure for (a) as-cast, (b) 400 $^{\circ}$C annealed, (c) 600 $^{\circ}$C annealed, (d) 800 $^{\circ}$C annealed, and (e) 1000 $^{\circ}$C annealed samples. (f) Temperature dependences of upper critical field of NbScTiZr samples. The figure is derived from (a)-(e). The solid curves show the fitting results using eq. (\ref{eq:hc2}).}
\end{center}
\end{figure}

We have also examined $\mu_{0}H_\mathrm{c2}$($T$) by measuring $\rho$($T$) under external fields, as depicted in Figs.\hspace{1mm}\ref{fig4}(a)-(e). 
The onset of the superconductive transition is employed as $T_\mathrm{c}$, and $\mu_{0}H_\mathrm{c2}$($T$) is again analyzed using the WHH model for each sample (see Fig.\hspace{1mm}\ref{fig4}(f)). 
Table \ref{tab:table2} also displays $\mu_{0}H_\mathrm{c2}$(0) and $\alpha_\mathrm{M}$ for each sample, denoted as $\mu_{0}H_\mathrm{c2}^{\rho}$(0) and $\alpha_\mathrm{M}^{\rho}$, respectively. 
The estimated values of $\lambda_\mathrm{SO}$ are 0.7, 0.3, 2.0, 4.5, and 4.5 for the as-cast, 400 $^{\circ}$C annealed, 600 $^{\circ}$C annealed, 800 $^{\circ}$C annealed, and 1000 $^{\circ}$C annealed samples, respectively. 
The dataset of $\mu_{0}H_\mathrm{c2}^{\rho}$(0) does not significantly differ from that of $\mu_{0}H_\mathrm{c2}^{M}$(0). 
Furthermore, the annealing temperature dependence of $\alpha_\mathrm{M}^{\rho}$ is similar to that of $\alpha_\mathrm{M}^{M}$.

\begin{table*}
\caption{\label{tab:table2}%
Superconducting parameters and Vickers microhardness of NbScTiZr. The superscripts $M$ and $\rho$ of $\mu_{0}H_\mathrm{c2}$(0), $\alpha_\mathrm{M}$, and $\xi_\mathrm{GL}$(0) denote the parameters evaluated through the temperature-dependent measurements of $M$ and $\rho$, respectively. $\lambda_\mathrm{GL}^{M}$(0) is calculated using $\mu_{0}H_\mathrm{c1}$(0) and $\xi_\mathrm{GL}^{M}$(0). $\lambda_\mathrm{GL}^{\rho}$(0) is calculated using $\mu_{0}H_\mathrm{c1}$(0) and $\xi_\mathrm{GL}^{\rho}$(0).   
$\kappa_\mathrm{GL}^{M}$ ($\kappa_\mathrm{GL}^{\rho}$) is calculated using $\xi_\mathrm{GL}^{M}$(0) ($\xi_\mathrm{GL}^{\rho}$(0)) and $\lambda_\mathrm{GL}^{M}$(0) ($\lambda_\mathrm{GL}^{\rho}$(0)).}
\begin{center}
\begin{tabular}{cccccc}
\hline
sample & as-cast & 400 $^{\circ}$C & 600 $^{\circ}$C & 800 $^{\circ}$C & 1000 $^{\circ}$C \\
\hline
$T_\mathrm{c}$ (K) & 8.11 & 8.71 & 9.36 & 9.13 & 9.03 \\
$\gamma_\mathrm{el}$ (mJ/mol$\cdot$K$^{2}$) & 6.47 & 6.79 & 5.77 & 5.46 & 6.62 \\
$\beta$ (mJ/mol$\cdot$K$^{4}$) & 0.150 & 0.127 & 0.127 & 0.129 & 0.145 \\
$\mu_{0}H_\mathrm{c1}$(0) (mT) & 29.1 & 32.9 & 22.9 & 22.2 & 4.46 \\
$\mu_{0}H_\mathrm{c2}^{M}$(0) (T) & 13.1 & 14.1 & 14.6 & 13.9 & 14.3 \\
$\alpha_\mathrm{M}^{M}$  & 1.7 & 1.8 & 1.4 & 1.3 & 1.3 \\
$\mu_{0}H_\mathrm{c2}^{\rho}$(0) (T) & 12.8 & 12.7 & 14.5 & 15.0 & 14.5 \\
$\alpha_\mathrm{M}^{\rho}$  & 1.7 & 1.9 & 1.4 & 1.3 & 1.3 \\
$\xi_\mathrm{GL}^{M}$(0) (nm) & 5.0 & 4.8 & 4.8 & 4.9 & 4.8  \\
$\lambda_\mathrm{GL}^{M}$(0) (nm) & 137 & 128 & 159 & 161 & 405 \\
$\kappa_\mathrm{GL}^{M}$ & 27 & 27 & 33 & 33 & 84 \\
$\xi_\mathrm{GL}^{\rho}$(0) (nm) & 5.1 & 5.1 & 4.8 & 4.7 & 4.8 \\
$\lambda_\mathrm{GL}^{\rho}$(0) (nm) & 137 & 127 & 159 & 162 & 405 \\
$\kappa_\mathrm{GL}^{\rho}$ & 27 & 25 & 33 & 34 & 84 \\
Hardness (HV) & 336(1) & 345(4) & 264(6) & 239(2) & 230(2) \\
\hline
\end{tabular}
\end{center}
\end{table*}

\begin{figure}
\begin{center}
\includegraphics[width=0.9\linewidth]{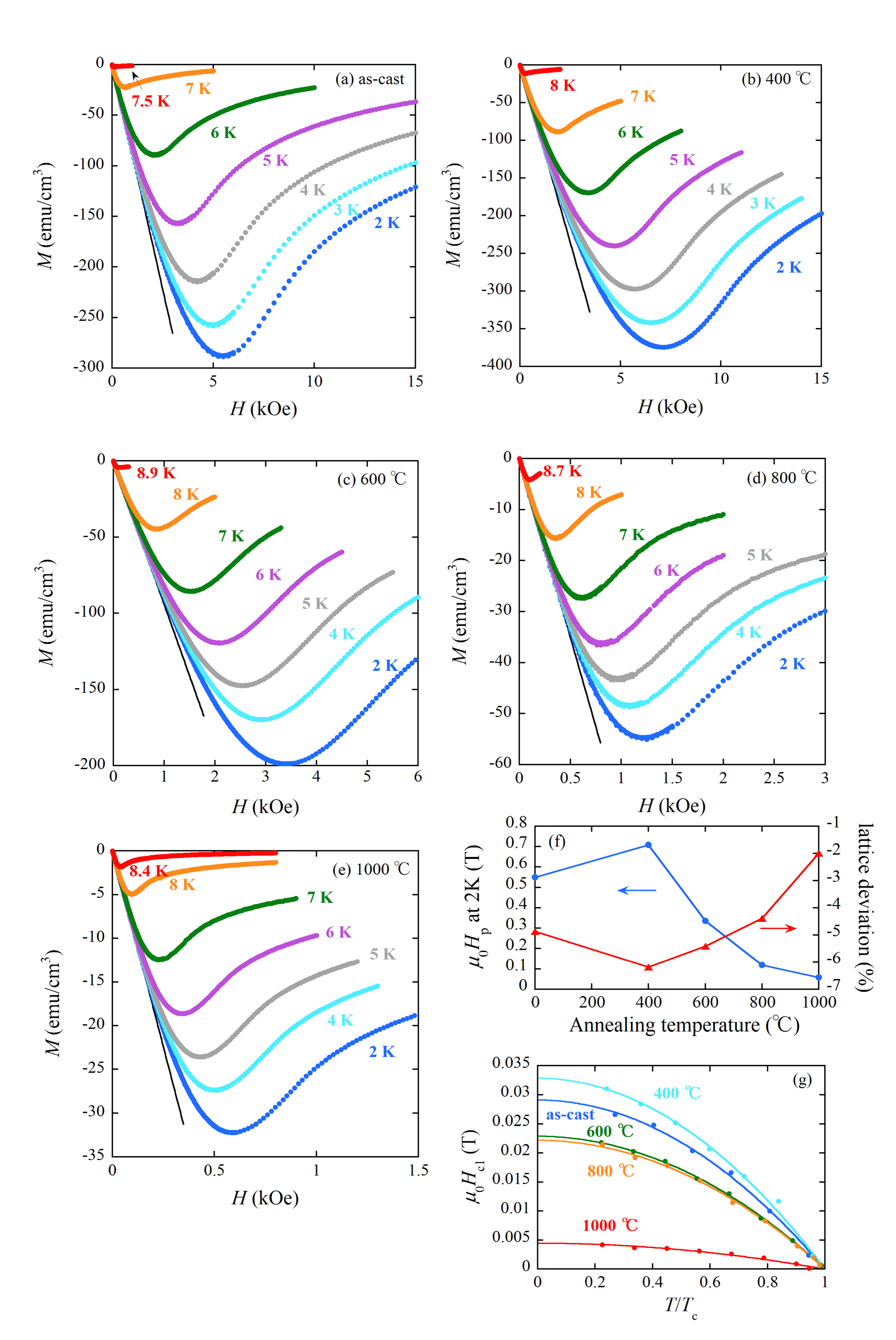}
\caption{\label{fig5}Field-dependent magnetization of (a) as-cast, (b) 400 $^{\circ}$C annealed, (c) 600 $^{\circ}$C annealed, (d) 800 $^{\circ}$C annealed, and (e) 1000 $^{\circ}$C annealed samples. (f) Annealing temperature dependences of $\mu_{0}H_\mathrm{p}$ at 2 K and lattice deviation. $\mu_{0}H_\mathrm{p}$ is defined as $\mu_{0}H$ showing the negative peak in the $M$-$H$ curve. (g) Temperature dependences of lower critical field of NbScTiZr samples. The solid curves show the fitting results using eq. (\ref{eq:hc1}).}
\end{center}
\end{figure}

Figures \ref{fig5}(a)-(e) exhibit the isothermal $M$-$H$ curves measured at various temperatures for all examined samples.
Notably, all samples exhibit negative broad peaks at $\mu_{0}H$ values, surpassing the anticipated lower critical field $\mu_{0}H_\mathrm{c1}$ for transition metal alloys. 
This phenomenon can generally be elucidated by strong flux pinning mechanisms\cite{Slimani:book}. 
In an ideal type-II superconductor with no defects, a completely reversible $M$-$H$ curve manifests, characterized by a sharp peak at $\mu_{0}H_\mathrm{c1}$. 
Introducing defects acting as flux pinning sites causes a sharp peak to be absent at the first flux quantum penetration, resulting in irreversibility in the $M$-$H$ curve. 
Consequently, the initial magnetization process displays a broad peak, as observed in Figs.\hspace{1mm}\ref{fig5}(a)-(e). 
It is worth noting that NbScTiZr in the as-cast state and heat-treated at 800 $^{\circ}$C exhibit extensive $M$-$H$ loops, as reported in the previous paper\cite{Seki:JSNM2023}. 
The area of the $M$-$H$ hysteresis loop for the as-cast sample exceeds that of the sample annealed at 800 $^{\circ}$C at the same temperature, indicative of robust flux pinning in the as-cast sample. 
Indeed, the external field demonstrating the negative peak ($\mu_{0}H_\mathrm{p}$) for the as-cast sample exceeds that of the 800 $^{\circ}$C annealed sample at the fixed temperature (refer to Figs.\hspace{1mm}\ref{fig5}(a) and (d)). 
The lattice deviation discussed in Fig.\hspace{1mm}\ref{fig1}(b) reflects lattice strain, correlating with flux pinning strength as the lattice strain yields pinning sites. Consequently, we examined the correlation between $\mu_{0}H_\mathrm{p}$ at 2 K and lattice deviation for each sample, as depicted in Fig.\hspace{1mm}\ref{fig5}(f). 
The figure demonstrates that a more significant lattice deviation corresponds to greater $\mu_{0}H_\mathrm{p}$ values. 
Therefore, flux pinning is most pronounced in the sample annealed at 400 $^{\circ}$C. 
In Figs.\hspace{1mm}\ref{fig5}(a)-(e), $\mu_{0}H_\mathrm{c1}$ value in each figure is determined by identifying the point at which the initial $M$-$H$ curve deviates from the linear line (see also solid line in each figure). 
The evaluated $\mu_{0}H_\mathrm{c1}$ increases with decreasing temperature, as demonstrated in  Fig.\hspace{1mm}\ref{fig5}(g), which is well reproduced by the Ginzburg-Landau equation, 
\begin{equation}
H_\mathrm{c1}(T)=H_\mathrm{c1}(0)\left(1-\left(\frac{T}{T_\mathrm{c}}\right)^{2}\right)
\label{eq:hc1}
\end{equation}
, where $\mu_{0}H_\mathrm{c1}$(0) represents the zero temperature $\mu_{0}H_\mathrm{c1}$. 
Table \ref{tab:table2} summarizes the values of $\mu_{0}H_\mathrm{c1}$(0) for all samples.

\begin{figure}
\begin{center}
\includegraphics[width=1\linewidth]{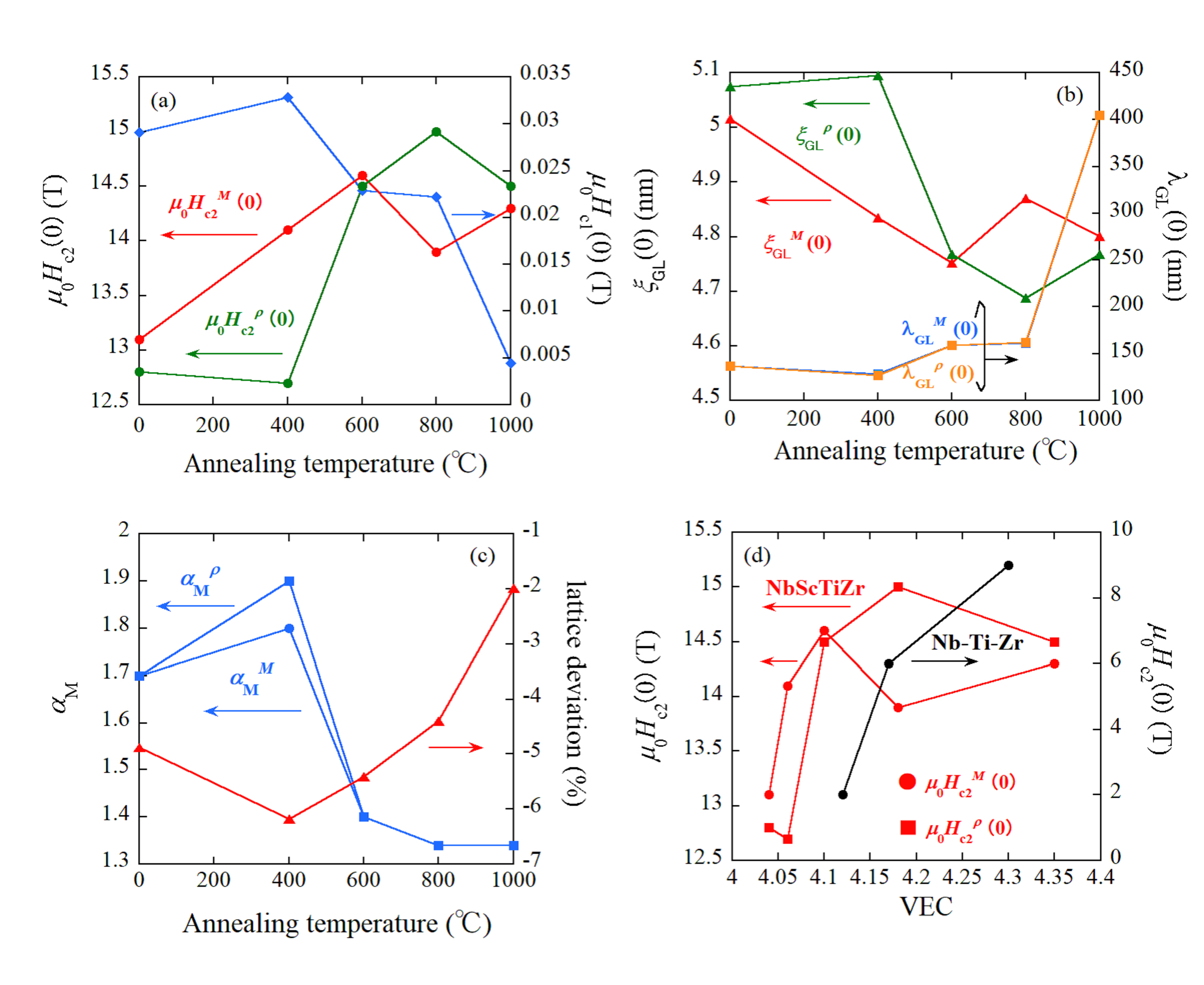}
\caption{\label{fig6} Annealing temperature dependences of (a) $\mu_{0}H_\mathrm{c1}$(0) and $\mu_{0}H_\mathrm{c2}$(0), (b) $\xi_\mathrm{GL}$(0) and $\lambda_\mathrm{GL}$(0), and (c) $\alpha_\mathrm{M}$. In (c), the annealing temperature dependence of lattice deviation is also shown for comparison. (d) VEC dependences of $\mu_{0}H_\mathrm{c2}$(0) for NbScTiZr and Nb-Ti-Zr alloys.}
\end{center}
\end{figure}

For each sample, the Ginzburg-Landau coherence length $\xi_\mathrm{GL}$(0) is calculated using the equation $\xi_\mathrm{GL}(0)=\sqrt{\frac{\Phi_{0}}{2\pi\mu_{0}H_\mathrm{c2}(0)}}$, where $\Phi_{0}$ denotes the magnetic flux quantum of 2.07$\times$10$^{-15}$ Wb. 
The resultant values are listed in Table \ref{tab:table2}. 
It is noted that $\xi_\mathrm{GL}$(0) is slightly smaller compared to typical HEAs (NbReHfZrTi: 6.1 nm\cite{Marik:JALCOM2018}, Ta$_{1/6}$Nb$_{2/6}$Hf$_{1/6}$Zr$_{1/6}$Ti$_{1/6}$: 5.2 nm\cite{Kim:AM2020}, and HfMoNbTiZr: 7.7 nm\cite{Kitagawa:JALCOM2022}). 
The magnetic penetration depth $\lambda_\mathrm{GL}$(0) can be extracted using the relation $\mu_{0}H_\mathrm{c1}(0)=\frac{\Phi_{0}}{4\pi\lambda_\mathrm{GL}(0)^{2}}\mathrm{ln}\frac{\lambda_\mathrm{GL}(0)}{\xi_\mathrm{GL}(0)}$. 
The extracted $\lambda_\mathrm{GL}$(0) values are also displayed in Table \ref{tab:table2}.
Subsequently, the Ginzburg-Landau parameter $\kappa_\mathrm{GL}$=$\lambda_{GL}(0)/\xi_{GL}(0)$ is obtained, yielding values ranging from 25 to 84, as shown in Table \ref{tab:table2}. 
All values exceed 1/$\sqrt{2}$, indicating type-II superconductors. 
Figures \ref{fig6}(a) and (b) encapsulate the annealing temperature dependences of fundamental superconducting parameters. 
The annealing temperature dependences of $\mu_{0}H_\mathrm{c2}$(0), $\lambda_\mathrm{GL}(0)$, and $\xi_\mathrm{GL}(0)$ do not exhibit systematic behaviors, indicating that flux pinning strength does not significantly alter the fundamental superconducting parameters. 
However, as shown in Fig.\ref{fig6} (c), the overall annealing temperature dependence of $\alpha_\mathrm{M}$ aligns well with that of lattice deviation. 
The value of $\alpha_\mathrm{M}$ estimated in each measurement reaches its maximum at an annealing temperature of 400 $^{\circ}$C, where the strongest flux pinning is observed.
In type-II superconductors, the Pauli paramagnetic field and the orbital limiting field are the principal sources of superconductivity disruption under a magnetic field. 
The critical magnetic field, which disrupts Cooper pairing due to the excess of Zeeman splitting energy over the superconducting gap energy, is termed the Pauli limit. 
In the orbital limiting field mechanism attributed to the Lorentz force, the field-induced kinetic energy of a Cooper pair surpasses the superconducting condensation energy.
The Maki parameter $\alpha_\mathrm{M}$ is given by $\frac{\sqrt{2}\mu_{0}H_\mathrm{c2}^\mathrm{orb}}{\mu_{0}H_\mathrm{c2}^\mathrm{Pauli}}$, where $\mu_{0}H_\mathrm{c2}^\mathrm{Pauli}$ is the Pauli limiting field and $\mu_{0}H_\mathrm{c2}^\mathrm{orb}$ is the orbital-limited upper critical field.
$\mu_{0}H_\mathrm{c2}^\mathrm{Pauli}$ is defined by 1.84$T_\mathrm{c}$. 
Employing refined $T_\mathrm{c}$ values listed in Table \ref{tab:table2}, $\mu_{0}H_\mathrm{c2}^\mathrm{Pauli}$ values are obtained as 14.9 T (as-cast), 16.0 T (400 $^{\circ}$C), 17.2 T (600 $^{\circ}$C), 16.8 T (800 $^{\circ}$C), and 16.6 T (1000 $^{\circ}$C). 
Substituting $\alpha_\mathrm{M}$ and $\mu_{0}H_\mathrm{c2}^\mathrm{Pauli}$ values into $\alpha_\mathrm{M}=\frac{\sqrt{2}\mu_{0}H_\mathrm{c2}^\mathrm{orb}}{\mu_{0}H_\mathrm{c2}^\mathrm{Pauli}}$, the datasets of $\mu_{0}H_\mathrm{c2}^\mathrm{orb}$ (T) are estimated as (17.9, 20.4, 17.0, 15.4, 15.3) obtained through $M$($T$) measurements and (17.9, 21.5, 17.0, 15.4, 15.3) obtained through $\rho$($T$) measurements for (as-cast, 400 $^{\circ}$C, 600 $^{\circ}$C, 800 $^{\circ}$C, 1000 $^{\circ}$C) samples.
In each dataset, $\mu_{0}H_\mathrm{c2}^\mathrm{orb}$ is the highest at an annealing temperature of 400 $^{\circ}$C, and the annealing temperature dependence of $\mu_{0}H_\mathrm{c2}^\mathrm{orb}$ roughly correlates with that of lattice deviation (also see Fig.\hspace{1mm}\ref{fig6}(c)). 
The larger lattice deviation seems to enhance $\mu_{0}H_\mathrm{c2}^\mathrm{orb}$.
As reported previously\cite{Seki:JSNM2023}, the lamellar structure exhibits fineness with a thickness of $\sim$ 70 nm in the as-cast state. 
Although the 400 $^{\circ}$C annealed sample maintains a fine microstructure, systematic grain size coarsening occurs with increasing annealing temperature above 400 $^{\circ}$C. 
It is anticipated that lattice strain is more pronounced in the fine eutectic structure due to the greater influence of the mismatch of structural properties at the interface of bcc and hcp phases, resulting in stronger flux pinning at lower annealing temperatures. 
Stronger flux pinning can elevate the orbital limiting field, which is consistent with our analysis of $\mu_{0}H_\mathrm{c2}^\mathrm{orb}$.

In the preceding discussion, we neglect the alteration of chemical composition upon annealing (refer to Table \ref{tab:table1}). 
The binary or ternary Nb-based bcc alloys adhere to the Matthias rule, which suggests that $T_\mathrm{c}$ is highly influenced by the density of states. 
Within transition metal bcc alloys, the density of states is delineated by the valence electron concentration (VEC), and $T_\mathrm{c}$ ascends with increasing VEC up to approximately 4.6\cite{Kitagawa:Metals2020}. 
The VEC values of bcc phases of NbScTiZr alloys are 4.04, 4.06, 4.10, 4.18, and 4.35 for the as-cast, 400 $^{\circ}$C annealed, 600 $^{\circ}$C annealed, 800 $^{\circ}$C annealed, and 1000 $^{\circ}$C annealed samples, respectively. 
Thus, as depicted in Fig.\hspace{1mm}\ref{fig3}(b), the trend of $T_\mathrm{c}$ versus annealing temperature conforms to the Matthias rule below 600 $^{\circ}$C. 
However, a departure from the Matthias rule is discerned above 600 $^{\circ}$C. 
To gain further insight into the composition effects on superconducting properties, the VEC dependence of $\mu_{0}H_\mathrm{c2}$(0) of NbScTiZr is juxtaposed with that of the Nb-Ti-Zr alloy, which bears resemblance to NbScTiZr, as shown in Fig.\hspace{1mm}\ref{fig6}(d). 
In the Nb-Ti-Zr system, (VEC, $\mu_{0}H_\mathrm{c2}$(0) (T)) datasets manifest as (4.12, 2) for Nb$_{0.12}$Ti$_{0.44}$Zr$_{0.44}$, (4.17, 6) for Nb$_{0.17}$Ti$_{0.62}$Zr$_{0.21}$, (4.3, 9) for Nb$_{0.30}$Ti$_{0.52}$Zr$_{0.18}$, and (4.3, 9) for Nb$_{0.33}$Ti$_{0.33}$Zr$_{0.33}$\cite{Ralls:JAP1980}.
Should the chemical composition comprehensively expound the annealing temperature dependence of $\mu_{0}H_\mathrm{c2}$(0), the VEC dependence of $\mu_{0}H_\mathrm{c2}$(0) in NbScTiZr would mirror that of Nb-Ti-Zr. 
However, a distinct behavior is observed. 
This suggests that the alteration of chemical composition does not sufficiently explain the annealing temperature dependence of the superconducting properties in NbScTiZr.

In contemplating the potential application as a superconducting wire, a comparison of $T_\mathrm{c}$ and $\mu_{0}H_\mathrm{c2}$(0) between NbScTiZr and NbTi alloy is imperative. 
The $T_\mathrm{c}$ and $\mu_{0}H_\mathrm{c2}$(0) of the NbTi alloy are 10 K and 14.5 T, respectively\cite{Slimani:book}.
Although most of the $\mu_{0}H_\mathrm{c2}$(0) values in NbScTiZr are comparable to that of the NbTi alloy, the $T_\mathrm{c}$ values of NbScTiZr are 1 to 2 K lower compared to the NbTi alloy. 
Hence, the pursuit of augmenting $T_\mathrm{c}$ without compromising $\mu_{0}H_\mathrm{c2}$(0) emerges as the subsequent paramount research endeavor.

\begin{figure}
\begin{center}
\includegraphics[width=1\linewidth]{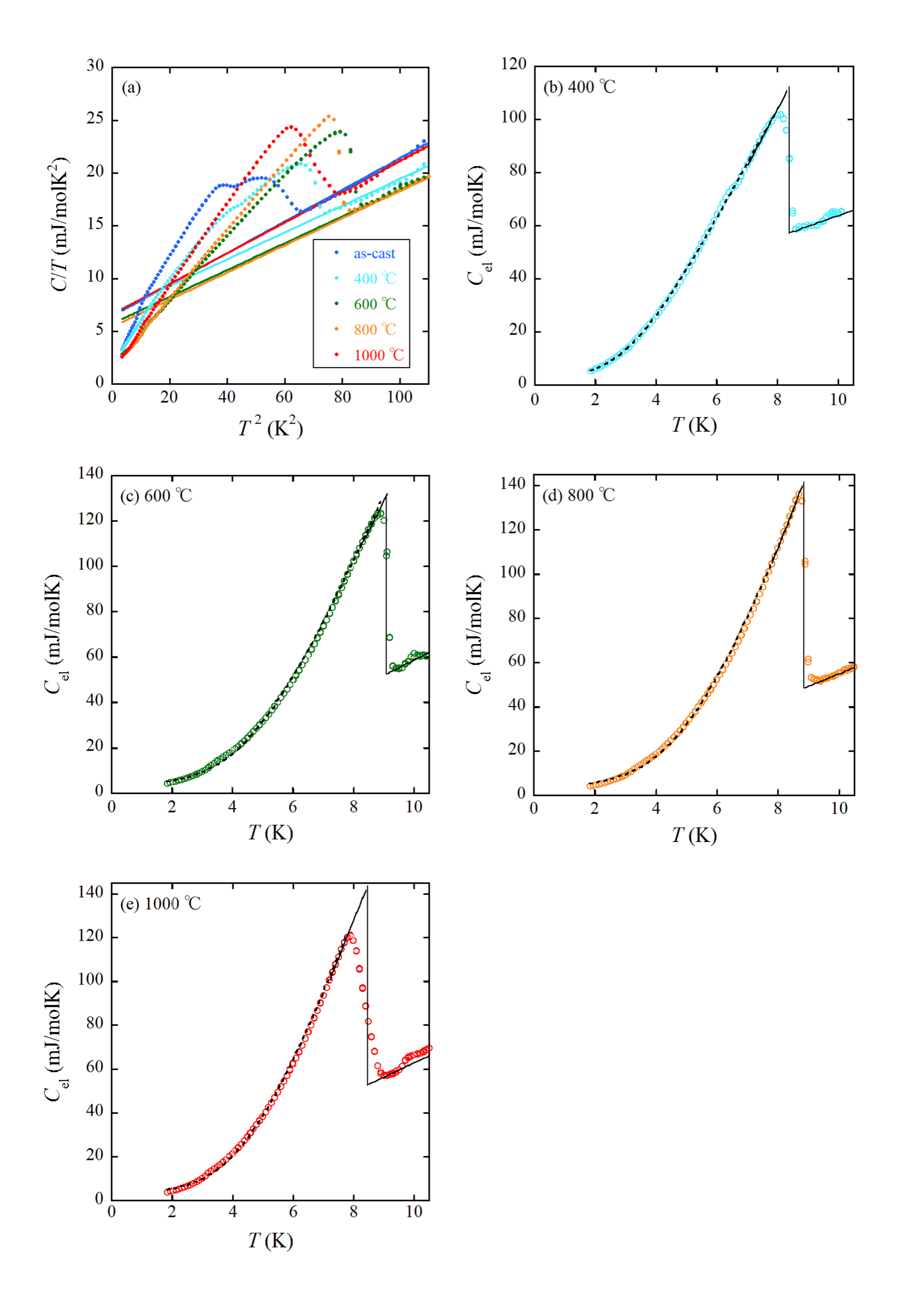}
\caption{\label{fig7}(a) $C/T$ vs. $T^{2}$ plots of NbScTiZr samples investigated. The solid lines represent the fitting results using $\frac{C}{T}=\gamma_\mathrm{el}+\beta T^{2}$ relation. Temperature-dependent $C_\mathrm{el}$ of (b) 400 $^{\circ}$C annealed, (c) 600 $^{\circ}$C annealed, (d) 800 $^{\circ}$C annealed, and (e) 1000 $^{\circ}$C annealed samples. The dotted lines show the fitting results using $A\mathrm{exp}\left(-\frac{\Delta(0)}{k_\mathrm{B}T}\right)+\gamma_\mathrm{el}^{'}T$. The solid lines serve as guides for extracting values of discontinuity in $C_\mathrm{el}$ at $T_\mathrm{c}$.}
\end{center}
\end{figure}

Figure \ref{fig7}(a) elucidates a comparison of the specific heat data for all samples, presented as $C/T$ vs. $T^{2}$ plots. 
In each sample, the conspicuous specific heat jump at $T_\mathrm{c}$ supports the bulk nature of the superconductive transition. 
Notably, the transition of the as-cast sample appears relatively broad, with an additional small anomaly detected at approximately 6.3 K.
The broad superconductive transition is observed in some bcc HEA superconductors. 
For instance, the single-phase bcc (NbTa)$_{0.67}$(MoHfW)$_{0.33}$, a 4.3 K $T_\mathrm{c}$ HEA superconductor, exhibits a broad specific heat jump, although the underlying cause of this broadening remains unclear\cite{Sobota:PRB2022}. 
Additionally, phase segregation occasionally yields a broadening of the transition, as seen in the case of bcc Ti-Hf-Nb-Ta-Re alloy represented by (Ti$_{20}$Hf$_{10}$)(Nb$_{35}$Ta$_{25}$)Re$_{10}$ and (Ti$_{15}$Hf$_{5}$)(Nb$_{35}$Ta$_{35}$)Re$_{10}$ \cite{Hattori:JAMS2023}. 
These alloys are composed of two bcc phases with slightly different chemical compositions. 
The specific heat jump of phase-segregated alloy tends to be broad due to an inhomogeneity of chemical composition. 
In the present case, the additional small anomaly below $T_\mathrm{c}$ may suggest the existence of an inhomogeneous distribution of chemical composition in the as-cast sample.
Therefore, the origin of the broad superconductive transition in the as-cast NbScTiZr alloy may be akin to that observed in the Ti-Hf-Nb-Ta-Re alloy. 
With increasing annealing temperature up to 800 $^{\circ}$C, the specific heat jump becomes sharper, but in the case of annealing at 1000 $^{\circ}$C, the specific heat jump exhibits slight broadening. 
As previously mentioned, the degree of lattice strain is most pronounced in the sample annealed at 400 $^{\circ}$C. 
The non-systematic behavior of the sharpness of the superconductive transition against the annealing temperature indicates the weak influence of lattice strain on the zero-field specific heat. 
Each data above $T_\mathrm{c}$ can be fitted by the equation 
\begin{equation}
\frac{C}{T}=\gamma_\mathrm{el}+\beta T^{2}
\label{eq:cp}
\end{equation}
, where $\gamma_\mathrm{el}$ and $\beta$ are the Sommerfeld coefficient and the lattice contribution, respectively. 
The resulting parameters are presented in Table \ref{tab:table2}. 
It is noteworthy that both $\gamma_\mathrm{el}$ and $\beta$ encompass contributions attributed to the bcc and hcp phases.

The superconducting parameters regarding the specific heat, except for the as-cast sample, are estimated through the analyses of Figs.\hspace{1mm}\ref{fig7}(b)-(e), which depict temperature dependences of the electronic specific heat $C_\mathrm{el}$($T$), obtained by subtracting $\beta T^{3}$ from $C$($T$). 
In the superconducting state, each $C_\mathrm{el}$($T$) can be modeled by $A\mathrm{exp}\left(-\frac{\Delta(0)}{k_\mathrm{B}T}\right)+\gamma_\mathrm{el}^{'}T$, where $A$ is a proportional constant, $\Delta(0)$ is the superconducting gap, $k_\mathrm{B}$ is the Boltzmann constant, and $\gamma_\mathrm{el}^{'}$ represents the electronic specific heat coefficient associated with the hcp phase. 
The determined values of $\Delta(0)$ (meV) and $\gamma_\mathrm{el}^{'}$ (mJ/mol$\cdot$K$^{2}$)) are (1.16, 2.8) for the 400 $^{\circ}$C annealed sample, (1.73, 2.7) for the 600 $^{\circ}$C annealed sample, (1.84, 2.8) for the 800 $^{\circ}$C annealed sample, and (1.66, 2.6) for the 1000 $^{\circ}$C annealed sample.
The $\gamma_\mathrm{el}^{'}$ values are necessary to achieve good fittings, as demonstrated by the dotted curves in Figs.\hspace{1mm}\ref{fig7}(b)-(e), indicating a non-superconducting behavior of the hcp phase down to 2 K. 
The values of $\frac{2\Delta(0)}{k_\mathrm{B}T_\mathrm{c}}$ are calculated as 3.22, 4.43, 4.85, and 4.56 for the 400 $^{\circ}$C, 600 $^{\circ}$C, 800 $^{\circ}$C, and 1000 $^{\circ}$C annealed samples, respectively. 
The value of $\frac{2\Delta(0)}{k_\mathrm{B}T_\mathrm{c}}$ for the 400 $^{\circ}$C annealed sample closely approximates 3.52, as expected for an s-wave BCS superconductor. 
Conversely, the remaining three values exceed 3.52, suggesting a possible strong-coupled superconductor. 
In the as-cast sample, the broadening of the transition impedes the evaluation of $\frac{2\Delta(0)}{k_\mathrm{B}T_\mathrm{c}}$. 
Next, we discuss the normalized specific heat jump $\frac{\Delta C_\mathrm{el}}{(\gamma_\mathrm{el}-\gamma_\mathrm{el}^{'}) T_\mathrm{c}}$, expected to be 1.43 for the BCS model in the weak coupling limit. 
In our analysis, we eliminate the contribution of $\gamma_\mathrm{el}^{'}$ from the $\gamma_\mathrm{el}$ obtained in Fig.\hspace{1mm}\ref{fig7}(a) because the hcp phase does not undergo a superconducting transition. 
$T_\mathrm{c}$ is employed as the midpoint of the transition. 
The discontinuities of $C_\mathrm{el}$ at $T_\mathrm{c}$ were estimated by extrapolating the data below and above $T_\mathrm{c}$, as depicted by solid lines in Figs.\hspace{1mm}\ref{fig7}(b)-(e). 
The estimated values of $\frac{\Delta C_\mathrm{el}}{(\gamma_\mathrm{el}-\gamma_\mathrm{el}^{'}) T_\mathrm{c}}$ are 1.63, 2.83, 3.97, and 2.68 for the 400 $^{\circ}$C, 600 $^{\circ}$C, 800 $^{\circ}$C, and 1000 $^{\circ}$C annealed samples, respectively. 
Similar to the case of $\frac{2\Delta(0)}{k_\mathrm{B}T_\mathrm{c}}$, the sample annealed above 600 $^{\circ}$C appears to largely deviate from the BCS superconductor in the weak coupling limit. 
It is worth noting that strong-coupled superconductivity has recently been reported in several transition metal-based bcc alloys, such as TiHfNbTa and Ta-Hf\cite{Li:JPCC2023,Meena:PRM2023}.
The sharp jump in specific heat at $T_\mathrm{c}$, particularly for the samples annealed at 600 $^{\circ}$C or 800 $^{\circ}$C, is intriguing, notwithstanding the fact that the volume fraction of the bcc phase approximates 50 \%. 
In these specimens, the $\gamma_\mathrm{el}^{'}$ value of the hcp phase parallels $\gamma_\mathrm{el}-\gamma_\mathrm{el}^{'}$ associated with the bcc phase. 
This implies that the contribution of specific heat from the hcp phase does not substantially surpass that of the bcc phase, thereby preventing the burying of the abrupt jump in specific heat of the bcc phase at $T_\mathrm{c}$.

We now address the annealing temperature dependence of $T_\mathrm{c}$. 
$T_\mathrm{c}$ generally depends on the density of states at the Fermi level and the electron-phonon interaction. 
The density of states at the Fermi level reflects the Sommerfeld coefficient of the bcc phase, represented by ($\gamma_\mathrm{el}-\gamma_\mathrm{el}^{'}$). 
The annealing temperature dependence of ($\gamma_\mathrm{el}-\gamma_\mathrm{el}^{'}$) does not align with that of $T_\mathrm{c}$ displayed in Fig.\hspace{1mm}\ref{fig3}(b). 
Therefore, the electron-phonon interaction would significantly influence $T_\mathrm{c}$. 
As depicted in Fig.\hspace{1mm}\ref{fig1}(b), thermal annealing has a notable impact on lattice strain, which would also alter the electron-phonon interaction, thereby affecting $T_\mathrm{c}$.

\begin{figure}
\begin{center}
\includegraphics[width=0.65\linewidth]{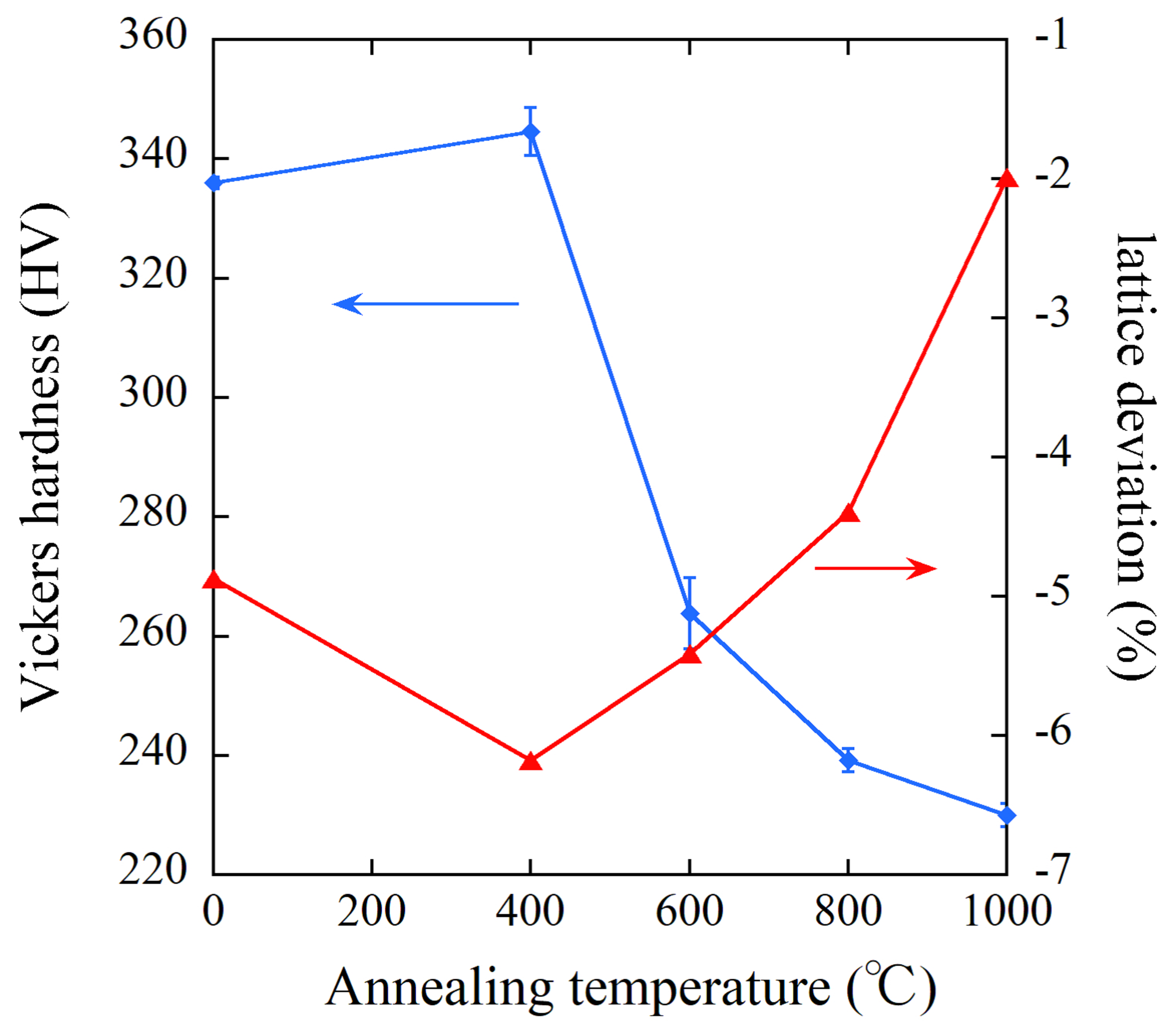}
\caption{\label{fig8}Annealing temperature dependence of Vickers microhardness for NbScTiZr. The annealing temperature dependence of lattice deviation is also shown for comparison.}
\end{center}
\end{figure}

Table \ref{tab:table2} also presents the Vickers microhardness values of the examined samples (see also Fig.\hspace{1mm}\ref{fig8}). 
Although annealing at 400 $^{\circ}$C leads to a slight enhancement of hardness compared to the as-cast state, increasing the annealing temperature above 400 $^{\circ}$C systematically diminishes the hardness. 
Lamellar structures are generally expected to coarsen at elevated annealing temperatures\cite{Fu:JMRT2022}. 
The microstructure of the as-cast NbScTiZr exhibits a significantly fine lamellar-like arrangement with a thickness of $\sim$ 70 nm. 
While the microstructure size remains consistent at 400 $^{\circ}$C annealing, grain size coarsening is observed as the annealing temperature exceeds 400 $^{\circ}$C\cite{Seki:JSNM2023}. 
A spheroidization-like morphology emerges in the sample annealed at 1000 $^{\circ}$C\cite{Seki:JSNM2023}. 
The microstructure-hardness relationship of NbScTiZr closely resembles that of CoCrFeNiNb$_{x}$ eutectic HEA\cite{He:MD2016}. 
The microstructure of CoCrFeNiNb$_{x}$ maintains a very fine lamellar arrangement up to  600 $^{\circ}$C annealing, beyond which the lamellar size systematically increases with rising annealing temperature. 
Above 600 $^{\circ}$C annealing, hardness shows a systematic decline. 
In the fine lamellar structure, interface strengthening is suggested as the origin of high hardness\cite{Xiong:JMST2021,Zhuang:Entropy2018}. 
The interface between two phases of the eutectic structure acts as a resistance to dislocation motion due to the mismatch of mechanical and structural properties at the interface. 
Consequently, the coarsened microstructure with reduced interface area results in decreased hardness. 
The microstructure-hardness relationship of NbScTiZr or CoCrFeNiNb$_{x}$ aligns well with the scenario of interface strengthening. 
Figure \ref{fig8} illustrates the annealing temperature dependence of Vickers microhardness alongside that of lattice deviation. 
As lattice deviation decreases, microhardness tends to decline. 
Lattice deviation reflects the degree of lattice strain, which is more pronounced in the fine eutectic structure due to the significant influence of structural property mismatches at the interface. 
Hence, the hardness-lattice deviation relation also supports the interface strengthening mechanism. 
Moreover, our results suggest that Vickers microhardness serves as an effective indicator of lattice strain in eutectic HEAs.

\section{Summary}
We investigated the superconducting properties of the as-cast and thermally annealed NbScTiZr samples, characterized by a eutectic structure comprising bcc and hcp phases, employing magnetic, transport, and zero-field thermal measurements. 
The lattice deviation, defined as the deviation of the experimental lattice parameter from that derived by the hard-sphere model with no lattice strain, suggests the greatest lattice strain at the annealing temperature of 400 $^{\circ}$C. 
We determined $\mu_{0}H_\mathrm{c1}$(0), $\mu_{0}H_\mathrm{c2}$(0), $\xi_\mathrm{GL}$(0), and $\lambda_\mathrm{GL}$(0) for all samples through magnetization and electrical resistivity measurements. 
The $M$-$H$ curves exhibiting negative peaks at $\mu_{0}H_\mathrm{p}$ much higher than $\mu_{0}H_\mathrm{c1}$ reveal stronger flux pinning for the as-cast, 400 $^{\circ}$C annealed, and 600 $^{\circ}$C annealed samples. 
Although $\mu_{0}H_\mathrm{c2}$(0), $\xi_\mathrm{GL}$(0), and $\lambda_\mathrm{GL}$(0) do not systematically depend on the annealing temperature, the annealing temperature dependence of the Maki parameter strongly correlates with that of lattice deviation.
Our analysis suggests that larger lattice strain can enhance the Maki parameter by elevating the orbital-limited field through the modification of flux pinning strength.
Specific heat measurements confirm the bulk nature of superconducting transition in all samples. 
However, the broadness of the superconductive transition in the as-cast state suggests the presence of chemical inhomogeneity. 
In the case of zero-field specific heat, the effect of lattice strain is less discernible compared to magnetization measurements. 
The possibility of strong-coupled superconductivity is supported by $\frac{2\Delta(0)}{k_\mathrm{B}T_\mathrm{c}}$ and $\frac{\Delta C_\mathrm{el}}{(\gamma_\mathrm{el}-\gamma_\mathrm{el}^{'}) T_\mathrm{c}}$, which exceed BCS values in the weak coupling limit. 
The annealing temperature dependence of Vickers microhardness correlates well with that of the lattice deviation. 
Our study indicates that the lattice parameter and Vickers microhardness effectively reflect the degree of lattice strain in eutectic HEAs. 
Additionally, lattice strain significantly influences the Maki parameter of eutectic HEA superconductors by modifying flux pinning strength.

\section*{Author Contributions}
Jiro Kitagawa: Conceptualization; Supervision; Formal analysis; Investigation; Writing – original draft; Writing – review \& editing.
Haruto Ueta: Formal analysis; Investigation.
Yuto Watanabe: Formal analysis; Investigation.
Takeru Seki: Formal analysis; Investigation.
Yoshikazu Mizuguchi: Formal analysis; Investigation; Writing - reviewing \& editing.
Terukazu Nishizaki: Formal analysis; Investigation; Writing - reviewing \& editing.

\section*{Declaration of Competing Interest}
The authors declare that they have no known competing financial interests or personal relationships that could have appeared to influence the work reported in this paper.

\section*{Acknowledgments}
J.K. acknowledges the support from a Grant-in-Aid for Scientific Research (KAKENHI) (Grant No. 23K04570) and the Comprehensive Research Organization of Fukuoka Institute of Technology. Y.M. acknowledges the support from a Grant-in-Aid for Scientific Research (KAKENHI) (Grant No. 21H00151). T.N. acknowledges the support from a Grant-in-Aid for Scientific Research (KAKENHI) (Grant No. 20K03867) and the Takahashi Industrial and Economic Research Foundation. We are grateful to Kousuke Ueda of Kyushu Sangyo University
for support of magnetization measurements.

%\section*{Declaration of Competing Interest}
%The authors have no conflicts to disclose.

\end{document}